\documentclass[pra,twocolumn,showpacs,a4paper,superscriptaddress]{revtex4}

\usepackage{dcolumn,graphicx}

\begin{document}

\title{Efficient and robust quantum key distribution %
       with minimal state tomography}


\author{Berthold-Georg~Englert}
\affiliation{Department of Physics, %
National University of Singapore, Singapore 117542, Singapore}
\affiliation{Centre for Quantum Technologies, %
National University of Singapore, Singapore 117543, Singapore}

\author{Dagomir~Kaszlikowski}
\affiliation{Department of Physics, %
National University of Singapore, Singapore 117542, Singapore}
\affiliation{Centre for Quantum Technologies, %
National University of Singapore, Singapore 117543, Singapore}

\author{Hui~Khoon~Ng}
\altaffiliation[Now at ]{Department of Physics, %
California Institute of Technology, Pasadena, CA 91125, USA}
\affiliation{Applied Physics Lab, DSO National Laboratories, %
Singapore 118230, Singapore}
\affiliation{Department of Physics, %
National University of Singapore, Singapore 117542, Singapore}

\author{Wee~Kang~Chua}
\affiliation{Applied Physics Lab, DSO National Laboratories, %
Singapore 118230, Singapore}
\affiliation{Department of Physics, %
National University of Singapore, Singapore 117542, Singapore}

\author{Jaroslav~\v{R}eh\'{a}\v{c}ek}
\affiliation{Department of Optics, Palacky University, 17.\ listopadu 50, 
772~00 Olomouc, Czech Republic}

\author{Janet Anders}
\affiliation{Department of Physics, %
National University of Singapore, Singapore 117542, Singapore}

\date{18 May 2008}

\begin{abstract}
We introduce the Singapore protocol, 
a qubit protocol for quantum key distribution that
is fully tomographic, more efficient than other tomographic protocols,
and very robust.
Under ideal circumstances the efficiency is $\log_2(4/3)=0.415$ key~bits
per qubit sent.
This is 25\% more than the efficiency of $1/3=0.333$
for the standard six-state protocol, which sets the benchmark.
We describe a simple two-way communication scheme that extracts $0.4$ key bits
per qubit and thus gets close to the information-theoretical limit.
The noise thresholds that we report for a
hierarchy of eavesdropping attacks demonstrate the robustness of the protocol:
A secure key can be extracted if there is less than 38.9\% noise. 
\end{abstract}


\begin{widetext}
\maketitle  
\end{widetext}

\section{Introduction}\label{sec:intro}
Almost all real-life implementations of schemes for quantum key distribution
use qubits as the carriers for the quantum information, and most of them
are variants of the very first proposal, the 1984 scheme by 
Bennett and Brassard (BB84, \cite{BB84}).
It has an efficiency of $1/2$ under ideal---that is,
noise-free---circumstances,
which is to say that, on average, one secure key bit can be extracted for two
qubits exchanged.

The measurements performed in BB84 do not explore all of the qubit's Bloch
sphere, but only probe a plane.
The tomography of the qubit state is, therefore, partial rather than complete,
which does not matter in ideal, noise-free operation, 
but it has its drawbacks when the quantum channel is noisy. 
By contrast, the ``six-state protocol'' (see, e.g., Ref.~\cite{6state}) offers
full state tomography; it is the qubit prototype of the higher-dimensional
tomographic protocols introduced in Ref.~\cite{TomoCrypt}.

The complete tomography comes at a price; 
for each key bit one has to exchange three qubits on average.
Full tomography can be had, however, at a much lower cost, 
as is demonstrated by the protocols with minimal qubit tomography (MQT) 
that we describe here. 
They have an ideal efficiency of $\log_2(4/3)=0.415$ 
and thus need only $2.41$ qubits per key bit.
Put differently, the MQT protocols are potentially 24.5\% more efficient than 
the six-state protocol, their competitor among the standard tomographic 
protocols.

The key observation is that the tomography of the six-state protocol is
redundant, inasmuch as one measures six probabilities to determine the three
parameters that specify the qubit state.
By contrast, one measures only four probabilities in MQT to
establish the three parameters. 
As discussed in Ref.~\cite{MQT}, qubit tomography of this minimal kind is
possible indeed.
Moreover, a simple one-loop interferometer setup can realize MQT
for the polarization qubit of a photon \cite{OneLoop}.
Non-interferometric setups are also possible \cite{NoLoop}, and work very well
in practice \cite{LPLLK1}.   

Key generation from the raw data is rather straightforward in the  
six-state protocol where one performs a basis matching as one does for BB84.
A similar procedure can be applied to the raw data of MQT protocols---the
``Renes pairing'' \cite{RenesTetra}, discussed in Sec.~\ref{sec:Renes}---but 
then the achieved efficiency is no more than $1/3=0.333$, the value of the
six-state protocol. 
Thus, if one wishes, as we do, to exploit the advantage offered by MQT,
one needs to abandon basis matching and the like in favor of another
procedure.
We introduce here one such alternative key generation method that is
simple to describe, and to implement, which extracts 0.4 key bits per qubit
and thus exceeds the benchmark value of $0.333$ by much and gets quite close 
to the information-theoretical limit of $0.415$. 

Here is a brief outline.
We begin with considering ideal circumstances in Sec.~\ref{sec:ideal}.
In Sec.~\ref{sec:2ways}, we then discuss the key generation by two-way
communication. 
The Singapore protocol is introduced in Sec.~\ref{sec:SingProt}, with 
emphasis on its tomographic element.  
In Sec.~\ref{sec:thresh}, we report the noise thresholds below which 
the Singapore protocol offers secure key distribution. 
We close with a summary.

The account given here is a general description of the scheme and a report of
all major results of the detailed analysis.
This analysis itself is presented in two companion 
papers~\cite{RawDataAtt,CohAtt}.

\section{Ideal circumstances}\label{sec:ideal}
Let us consider ideal circumstances for a start.
We emphasize the symmetry between the communicating parties---Alice and
Bob---by a scenario in the spirit of Ekert's entanglement-based
protocol of 1991 \cite{E91}.
A source distributes entangled qubits to Alice and Bob, one pair at a time.
Ideally, the source emits the pairs in the singlet state $\ket{s}$, 
whose statistical operator
\begin{equation}
  \label{eq:singlet}
\rho_\mathrm{AB}=\ket{s}\bra{s}=\frac{1}{4}
\bigl(1-\vec{\sigma}_\mathrm{A}\cdot\vec{\sigma}_\mathrm{B}\bigr)
\end{equation}
is a symmetric function of the Pauli vector operators 
$\vec{\sigma}_\mathrm{A}$ and $\vec{\sigma}_\mathrm{B}$
for Alice's and Bob's qubit, respectively.

For each qubit pair, one of the four detectors at Alice's end will respond, 
and one of Bob's four detectors. 
Each set of detectors realizes, for the respective qubit, 
the probability operator measurement (POM---in the mathematical literature:
POVM for positive operator valued measure)
that consists of the four half-projectors
\begin{equation}
  \label{eq:POM}
  P_k = \frac{1}{4} \bigl( 1 + \vec{t}_k \cdot \vec{\sigma}\bigr)\,, 
\quad\mbox{for}\ k =1,\dots, 4\,,
\end{equation}
where the 
unit vectors $\vec{t}_k$ have equal angles between them,
\begin{equation}
  \label{eq:tetra}
  \vec{t}_k\cdot\vec{t}_l=\frac{4}{3}\delta_{kl}-\frac{1}{3}\,,
\quad\mbox{for}\ k,l =1,\dots, 4\,.
\end{equation}
Geometrically speaking, they are the normal vectors for the four faces 
of a tetrahedron, or one can picture them as pointing from the center of a
cube to four nonadjacent corners; see Fig.~1 in Ref.~\cite{MQT} for an
illustration.
Their linear dependence and completeness, stated respectively by
\begin{equation}
  \label{eq:basictetra}
  \sum_{k=1}^4\vec{t}_k=0\quad\mbox{and}\quad
  \frac{3}{4}\sum_{k=1}^4\vec{t}_k\vec{t}_k=\tensor{1}\,,
\end{equation}
are basic properties of the tetrahedron vector quartet.

With $\vec{\sigma}\to\vec{\sigma}_\mathrm{A}$ in \Eqref{POM} we have the
members $P_{\mathrm{A}k}$ for Alice's POM, and likewise
$\vec{\sigma}\to\vec{\sigma}_\mathrm{B}$ gives Bob's POM.
The 16 joint probabilities that Alice's $k$th detector fires together with
Bob's $l$th detector are then given by
\begin{eqnarray}
  \label{eq:jointprobs}
  p_{kl}&=&\expect{P_{\mathrm{A}k}P_{\mathrm{B}l}}
\nonumber\\&=&
\frac{1}{16}\tr{\rho_\mathrm{AB} 
\bigl( 1 + \vec{t}_k \cdot \vec{\sigma}_\mathrm{A}\bigr)
\bigl( 1 + \vec{t}_l \cdot \vec{\sigma}_\mathrm{B}\bigr)}
\end{eqnarray}
for $k,l=1,2,3,4$.
Knowledge of these $p_{kl}$s is tantamount to knowing $\rho_\mathrm{AB}$
because the inverse relation \cite{MQT}
\begin{equation}
  \label{eq:invert}
    \rho_\mathrm{AB} = \sum_{k,l=1}^{4} 
    \bigl(6 P_{\mathrm{A}k}-1\bigr)p_{kl}\bigl(6 P_{\mathrm{B}l}-1\bigr)  
\end{equation}
reconstructs $\rho_\mathrm{AB}$ from the joint probabilities.
Indeed, two-qubit tomography of this kind can be performed and is highly
reliable in practice \cite{LPLLK2}.

We can think of Alice's measurement as preparing Bob's qubit at random in the
state corresponding to---that is, conditioned on---her unpredictable
measurement result. 
In this sense, she is sending him a random sequence of qubits through an
effective quantum channel. 
In fact, if the pair source is located in Alice's laboratory, the pair
emission in conjunction with her measurement is fully equivalent to a
single-qubit source.    
Conversely, if Alice is really operating a single-qubit source with random
output, so that the scenario is that of BB84, we can think of it 
as an Ekert scenario with a corresponding effective two-qubit source.
Therefore, our analysis applies to either physical situation.

For the ideal source that emits the singlet states of \Eqref{singlet}, we have 
\begin{equation}
  \label{eq:pkl-ideal}
  p_{kl}=\frac{1-\delta_{kl}}{12}=\left\{
    \begin{array}{c@{\mbox{\ for\ }}l}
    0 & k=l\,, \\ 1/12 & k\neq l\,,
    \end{array}\right.
\end{equation}
that is, two corresponding detectors ($k=l$) never fire together and the other
twelve cases are equally probable.
Accordingly, the mutual information between Alice and Bob is
\begin{equation}
  \label{eq:IABideal}
  I_\mathrm{AB}=\sum_{k,l =1}^{4} p_{kl} \log_2
        \frac{p_{kl}}
         {p_{k\cdot}p_{\cdot l}}
        = \log_2\frac{4}{3}= 0.415\ \mbox{(bits)},
\end{equation}
where 
\begin{equation}
  \label{eq:marginals}
p_{k\cdot}=\expect{P_{\mathrm{A}k}}=\sum_{l=1}^4 p_{kl}\,,\quad
p_{\cdot l}=\expect{P_{\mathrm{B}l}}=\sum_{k=1}^4 p_{kl}  
\end{equation}
are the marginal probabilities for her and him, respectively; here simply 
$p_{k\cdot}=p_{\cdot l}=\frac{1}{4}$.
This $I_\mathrm{AB}$ value exceeds the value of
$\frac{1}{3}$ for the six-state protocol \cite{6IAB} by almost 25\%.

The number $I_\mathrm{AB}=0.415$ tells us that Alice and Bob can generate up to
$0.415$ secure key bits for every qubit exchanged through the quantum 
channel, that is, for every qubit pair they detect.
If they had an appropriate error-correcting code at hand, the key generation
could be done by \emph{one-way} communication. 
Unfortunately, however, the best codes that are presently available have an
efficiency below the $0.333$ benchmark set by the six-state protocol
\cite{CodingGroup,benchmark}.

\section{Key generation by two-way communication}\label{sec:2ways}
The standard methods for generating the secret key-bit sequence---for the
BB84 scheme, the six-state protocol, and others---all make use of 
\emph{two-way} communication. 
Such procedures can also be designed for raw data characterized by the joint
probabilities in \Eqref{pkl-ideal}.

For ease of presentation, we assign letters A, B, C, D to the measurement
results $k,l=1,2,3,4$, so that the raw data make up 
a four-letter random sequence for Alice and another one for Bob. 
These sequences are such that the two letters for a qubit pair are never the
same, and the twelve pairs of different letters are equally frequent.
It is \emph{as if} Alice were sending a random sequence of the four letters to
Bob, who never receives the letter sent but gets either one of the other three
letters equally likely.

\subsection{Renes pairing}\label{sec:Renes}
Renes's method for key generation \cite{RenesTetra} amounts to the following.
Suppose Alice has letter A, to which she assigns $0$ or $1$ at random, say $1$
to be specific.
Then she chooses randomly one of the other three letters, say B, and
communicates publicly to Bob: 
``If your letter is A, call it $0$; if you have B, call it $1$!'', whereby the 
letter for value $0$ is always stated first. 
Bob in turn reports success if his letter is A or B, failure if he has C or D.

Since Alice has a $1/3$ chance of guessing Bob's letter right when she pairs B
with A, the success probability is $1/3$ and the failure probability is $2/3$.
In case of success, a pair of potential outcomes is identified with perfect
\mbox{(anti-)}\-cor\-re\-la\-tions and a key bit is obtained, in full analogy
to the pairs of outcomes that are selected by the basis matching of the BB84
protocol or the six-state protocol.

Indeed, Renes's method is in the tradition of basis-matching procedures
inasmuch as his pairing selects a $2\times2$ submatrix from the $4\times4$
matrix of the joint probabilities of \Eqref{pkl-ideal}. 
There is a crucial difference though: 
Only certain submatrices are useful in the BB84 protocol and the six-state
protocol and they are systematically chosen by the basis matching.
But for the MQT probabilities of \Eqref{pkl-ideal}, the choice is not unique;
you can pair A with B or with C or with D.
Possibly useful correlations are unavoidably discarded by the Renes pairing
in the unlucky failure cases and, therefore, this method does not take good
advantage of the stronger MQT correlations and does not yield a
higher efficiency than the six-state protocol.

\subsection{Beyond pairing: Iterative key extraction}\label{sec:iteration}
One can do much better by other procedures that do not rely on variants of
BB84 basis matching and the like. 
We describe here one such method, which generates the secret key bits by an
iteration and is rather simple to implement. 

We recall that the raw data consist of a four-letter random sequence
for Alice and another one for Bob.
Each round of the iterative extraction of the key bits then involves the
following steps. 

\textbf{Step~1:}~Alice chooses one letter at random, say A. 
She announces two positions in her sequence where this letter occurs, 
while not revealing which letter she has chosen.

\textbf{Step~2a:}~\emph{If Bob has two different letters at these 
positions}, say C and D, he forms two groups, one consisting of his letters
(CD), one of the others (AB). 
He knows for sure that Alice's letter is in the second group.
He decides at random to which group he assigns value $0$ and to 
which value $1$.
Then he announces the two groups and their values. 
Alice and Bob both enter the value of the group that contains Alice's letter
as the next bit of the secret key.

\textbf{Step~2b:}~\emph{If Bob has the same letter twice},
he announces that this is the case, not telling, of course, which
letter he got. 
Alice records her letter as part of a new sequence for later use. 
Bob does the same with his letter.
The secondary sequences thus formed have the same statistical properties as
the primary sequences.  

Alice and Bob repeat steps 1 and 2 as long as there are enough unused letters.
Then they apply the same two-step procedure to the new sequences created in
step~2b, thereby getting more key bits and two other new sequences. 
Next, these sequences are processed, and so forth.

As a result, Alice and Bob will share an identical sequence of key bits.
Nobody else knows the key bits because the public announcements
by Alice in step~1 and by Bob in steps~2a or 2b reveal nothing at all about
the values of the key bits.

When a key bit is generated from the original sequences, which happens with
probability $\frac{2}{3}$, two letters are consumed to get it.
For a key bit from the first put-aside sequences, the probability is
$\frac{1}{3}\times\frac{2}{3}$ and four letters are used altogether.
For a key bit from the second put-aside sequences, eight letters are spent
with a success probability of $\frac{1}{3}\times\frac{1}{3}\times\frac{2}{3}$;
and so forth. 

If the original sequences are $N$ letters long, we thus get 
$(\frac{1}{3}+\frac{1}{18}+\frac{1}{108}+\cdots)N$ key bits in total. 
The asymptotic efficiency is therefore
\begin{equation}
  \label{eq:asymeff}
  \lim_{n\to \infty}\frac{2}{5}\left[
    1-\left(\frac{1}{6}\right)^{\!n\,}\right] =\frac{2}{5} = 0.4\,,
\end{equation}
which falls short of the theoretical maximum of $0.415$, 
but not by much \cite{GetAll}.
Owing to the geometric convergence, just a few rounds are sufficient in
practice, the efficiencies being $0.333$, $0.389$, $0.398$ for one, two,
three rounds, respectively.
In short, it is easy to get above the $\frac{1}{3}$ efficiency of the six-state
protocol.

\subsection{Hybridization}\label{sec:hybrid}
In any practical implementation of the iterative key generation one has to
settle for a finite number of rounds. 
It is then beneficiary to apply the Renes pairing in the \emph{final} iteration
round, i.e.,
\textbf{Step~2b':}~\emph{If Bob has the same letter twice},
he chooses another letter at random to form a Renes pair, which he
communicates to Alice.

The efficiency is then
\begin{equation}
  \label{eq:asymeff'}
  \frac{2}{5}\left[1-\left(\frac{1}{6}\right)^{\!n\,}\right] 
  +\frac{1}{3}\left(\frac{1}{6}\right)^{\!n\,}=
  \frac{2}{5}\left[1-\left(\frac{1}{6}\right)^{\!n+1\,}\right]
\end{equation}
for a total number of $n$ rounds.
Thus, this hybrid reaches the $n$-iteration yield of \Eqref{asymeff} after
$n-1$ rounds.

\section{The Singapore protocol}\label{sec:SingProt}
The tetrahedron version of Renes's ``spherical codes'' \cite{RenesTetra} uses
the pairing method of Sec.~\ref{sec:Renes} for the key generation from the raw
data obtained by MQT.
It does not take advantage of the tomographic power offered by MQT.

In marked contrast, systematic state tomography is a defining element of the
\emph{Singapore protocol} on which we will focus now.
The Singapore protocol is specified by (i) MQT for the acquisition of raw
data; (ii) state tomography for the characterization of the source; (iii) key
generation by the iteration method of Sec.~\ref{sec:iteration}, or by the
hybrid method of Sec.~\ref{sec:hybrid}, or by some other procedure of high
efficiency.
The choice of method in (iii) distinguishes variants of the Singapore
protocol, but they all have (i) and (ii) in common.

\subsection{The tomographic element}\label{sec:tomography}
The tomographic element in the Singapore protocol exploits \Eqref{invert}.
After the detection of many qubits, Alice and Bob select a random subset of
the qubit pairs and reveal to each other, and everybody else, the measurement 
results obtained for them.
The relative frequencies of the pair events---for which fraction of the pairs
did she get, say, a B and he a D?---can serve as simple estimates for the
joint probabilities ($p_{24}$ for the BD events). 
When used in \Eqref{invert}, they provide an estimate of the two-qubit state
emitted by the source.
The more refined methods of quantum state estimation \cite{QSE} are likely to
give better and more reliable estimates, if the need arises, but in the context
of the Singapore protocol this is less important. 
 
Rather, Alice and Bob check whether the actual relative frequencies are
consistent with the expected output of the qubit-pair source.
They proceed with the key generation only if the source passes the test.
If it fails, the data are rejected as unreliable, and a different 
source is used.
The ``source'' encompasses here everything that is involved in delivering
the qubits to Alice and Bob: the physical source plus the transmission lines.
In the equivalent scenario in which Alice produces qubits and sends them to
Bob, the tomography establishes the properties of the quantum
channel through which the qubits are transmitted. 
One can then judge whether the channel is acceptable or not.

One could think that an essential part of the checking is to make sure that
the qubit pairs are statistically independent.
This would require a careful look at the joint probabilities for two or more
qubit pairs.
With a limited number of data sacrificed for the purpose of
tomography, the confidence level will be high for two pairs, 
lower for three pairs, even lower for four pairs, and so forth.
But if Alice and Bob are willing to pay the price, they can reach the
confidence level they desire.    

In fact, however, such elaborate checks of statistical independence are not
necessary because one can rely on the quantum version of the de Finetti theorem
\cite{Renner}.
It ensures that the ensemble of qubit pairs is, for all practical purposes,
equivalent to an ensemble of statistically independent pairs if all operations
(consistency check of the relative frequencies of paired qubits iterative key
generation) use randomly selected pairs.

It is worth emphasizing that Alice and Bob must have criteria according to
which they decide whether the source is trustworthy.
With a finite number of data at hand, there can never be absolute certainty
about statistical properties. 
In this respect, the tomography of the Singapore protocol, and the conclusions
about security drawn from it, is on equal footing with the typicality
assumptions in the security analysis of schemes for quantum key distribution
that follow the paradigm of BB84. 

It is, so to say, a matter of your level of distrust. 
When testing a coin that is supposedly unbiased, many would have serious doubts
if a million trials gave 70\% heads. 
But if you are leery, you might mistrust the coin after getting
seven heads in ten trials.

\subsection{Acceptable sources}\label{sec:accept}
An acceptable source would emit independent qubit pairs in the singlet state 
of \Eqref{singlet}
with an admixture of unbiased noise---or, equivalently, the noise of an
acceptable transmission channel must be unbiased \cite{OtherSources}.
Accordingly, the statistical properties of the observed detection events must
be consistent with the joint probabilities
\begin{equation}
  \label{eq:noisy_jointprobs}
    p_{kl} = \frac{4-\epsilon}{48}(1 - \delta_{kl})              
             + \frac{\epsilon}{16}\delta_{kl} 
\end{equation}
that derive from the two-qubit state of the form
\begin{equation}
  \label{eq:noisysinglet}
  \rho_\mathrm{AB}=\ket{s}(1-\epsilon)\bra{s}+\frac{\epsilon}{4}\,,
\end{equation}
where $\epsilon$ specifies the noise level: no noise for $\epsilon=0$; nothing
but noise for $\epsilon=1$. 
The range of interest is $0\leq\epsilon<\frac{2}{3}$ because 
$\rho_\mathrm{AB}$ is separable if $\epsilon\geq\frac{2}{3}$, and then
the correlations exhibited by the $p_{kl}$s are of a classical nature and
possess no genuine quantum properties.
The mutual information between Alice and Bob for the $p_{kl}$s of
Eq.~(\ref{eq:noisy_jointprobs}), 
\begin{equation}\label{eq:Iab}
    I_\mathrm{AB} = \left(1 - 
    \frac{\epsilon}{4}\right)  
    \log_2\left(\frac{4 - \epsilon}{3}\right) + 
    \frac{\epsilon}{4}\log_2\epsilon\,,
\end{equation}   
decreases monotonically from the $\epsilon=0$ value of $0.415$
to $0.0292$ for $\epsilon=\frac{2}{3}$. 

It is improbable that the noise in real quantum channels is truly unbiased.
Therefore, Alice and Bob ensure the joint probabilities of
Eq.~(\ref{eq:noisy_jointprobs}) by twirling:
for each pair of detection events the assigning of the letters A, B, C, D to
the detectors is done by a random choice of one of the 24 permutations of the
letters.

\subsection{Eavesdropping}\label{sec:eaves}
All noise is potentially resulting from eavesdropper Eve's attempts at
acquiring knowledge about the key bits.
She is given complete control over the source, and the best she can do is to
prepare a pure state in which all the qubit pairs sent to Alice and Bob are
entangled with a gigantic ancilla.

Upon tracing over the ancilla degrees of freedom, we get the state of many
qubit pairs received by Alice and Bob which---as a result of the full
tomography or as an implication of the quantum de Finetti theorem---is known
to be a product state with one factor for each qubit pair. 
As a consequence of the Schmidt decomposition of the pure entangled state, the
reduced ancilla state is unitarily equivalent to this product state, so that
Eve's ancilla consists of independent qubit pairs, one for each pair sent to
Alice and Bob.

Accordingly, Eve's optimal strategy \cite{RawDataAtt} amounts to entangling
each qubit pair emitted by the source with a two-qubit ancilla, and to keep
all these ancillas as quantum records when the qubit pairs are sent to Alice
and Bob.  
Eve must ensure that the noisy singlet state
\refeq{noisysinglet} results when $\ket{S}\bra{S}$ is traced over the
ancilla degrees of freedom, where $\ket{S}$ is the ket for the joint state of
a qubit pair sent to Alice and Bob and the qubit pair of its ancilla. 
This requirement determines $\ket{S}$ completely, 
up to irrelevant unitary transformations on the ancilla (see \cite{TomoCrypt}
and the appendix in \cite{pyramids}), namely,
\begin{equation}
  \label{eq:SourceState}
  \ket{S}=\ket{s_{12}s_{34}}\sqrt{1-\epsilon}
          +\ket{s_{13}s_{24}}i\sqrt{\epsilon}\,.
\end{equation}
Here, $\ket{s_{jk}}$ is the singlet for qubits $j$ and $k$, whereby qubits $1$
and $2$ are sent to Alice and Bob, and qubits $3$ and $4$ compose the ancilla.
Thus, in this most symmetric choice for $\ket{S}$, the qubit pair of Eve's
ancilla is on equal footing with the pair sent to Alice and Bob.

\section{Noise thresholds for the Singapore protocol}\label{sec:thresh}
We consider four different eavesdropping attacks.
In the first attack, Eve tries to learn as much as she can about the original
letter sequences recorded by Alice and Bob and does not wait until the public
communication between them reveals additional clues.
Eve can thus measure her ancillas immediately after the qubits haven been sent
to Alice and Bob, or even before the qubits leave the source, so that this
attack could be realized by just emitting a sequence of mixed two-qubit states
from the source.
This \emph{mixed-state attack} is relevant if Eve cannot store the
ancillas for later processing and Alice and Bob use a suitable code, 
such as those discussed in Ref.~\cite{CodingGroup}, to generate
the cryptographic key by one-way communication.
If Eve can, however, process the ancillas as late as she wishes, we have the
situation discussed in Sec.~\ref{sec:message1way}.

In the second attack, the \emph{raw-data attack} on the generated key bits,
Eve takes into account what she learns from the public communication between
Alice and Bob when they execute the steps of Secs.~\ref{sec:iteration} and
\ref{sec:hybrid}. But she continues to rely on the data she gains by the
mixed-state attack. 
Given the limitations of present-day technology, this is the strongest attack
Alice and Bob have to fear in practice.

A much stronger attack, however, is the third attack we
consider, the \emph{collective attack} on the generated key bits.
Here, Eve performs a collective measurement on all the ancillas to those qubit
pairs that contribute to the key bit under attack, exploiting fully the
information revealed publicly by Alice and Bob while they execute steps of
Secs.~\ref{sec:iteration} and \ref{sec:hybrid}.  
Depending on the iteration round in which the key bit is generated, this will
involve two, four, eight, \dots\ ancillas, each of them itself a qubit pair.

In order to implement the collective attack, Eve must be able to store the 
ancillas for a long time, because Alice and Bob can process their raw data 
long after it has been acquired, and she must be able to process the ancillas 
jointly.
Both tasks are beyond the quantum technology of today and the foreseeable
future. 
But once the technology exists, such attacks will be a matter of real concern.

Therefore, we should also consider the strongest attack conceivable: the
\emph{message attack}.
Here, Eve waits even longer, until after Alice and Bob have processed the keys
generated by the steps of Secs.~\ref{sec:iteration} and \ref{sec:hybrid} by
the standard methods of privacy amplification and error correction, and have
used the final sequence of key bits to encrypt a message. 
Rather than trying to learn the values of the key bits at any intermediate
stage, Eve attempts to decrypt the message itself, thereby making optimal use
of \emph{all} the public information exchanged between Alice and Bob up to,
and including, the encrypted message.

We are content here with giving brief accounts of the results of analyzing
these eavesdropping attacks.  
The technical details are reported in two companion papers,
Refs.~\cite{RawDataAtt} and \cite{CohAtt}.

\subsection{Mixed-state attack}\label{sec:raw}
For each of Alice's detection events, there is a conditioned ancilla state of
rank~2.
It is as if Alice were sending a random sequence of four different ancilla
states to Eve, each occurring one-fourth of the time.
Eve extracts the information accessible to her by measuring the
ancillas with the POM that is optimal for this purpose.
For $\epsilon>0.1725$,
Eve's optimal POM has four possible outcomes and its structure is 
analogous to the optimal POM for the six-state protocol that is given in 
Ref.~\cite{AccInf}.
For  $\epsilon<0.1725$, the optimal POM has five elements and extracts
slightly more information (less than 1\%) than the four-member POM.

In the more relevant range of $\epsilon>0.1725$, the joint probabilities for
Alice's $k$th result and Eve's $l$th result are given by the $p_{kl}$s of
\Eqref{jointprobs} with $\epsilon$ replaced by 
\begin{equation}
  \label{eq:eps2eta} \textstyle
  \eta=\Bigl(\sqrt{1-\frac{3}{4}\epsilon}
       -\sqrt{\frac{3}{4}\epsilon}\,\Bigr)^2\,.
\end{equation}
Upon presenting this relation in the form
\begin{equation}
  \label{eq:eps+eta}
\textstyle  (1-\frac{3}{2}\epsilon)^2+(1-\eta)^2=1\,,
\end{equation}
we observe a fundamental symmetry between the noise level $\epsilon$ in the
Alice--Bob channel and the noise level $\eta$ in the Alice--Eve channel. 
Note in particular that $\eta=1$ for $\epsilon=0$ and $\eta=0$ 
for $\epsilon=\frac{2}{3}$.
That is, when there is no noise in the channel from Alice to Bob, there is
nothing but noise between Alice and Eve; and when Alice and Bob receive their
qubits in a separable state, the effective transmission from Alice to Eve is
noiseless.

\begin{figure}[t]
\centerline{\includegraphics[bb=80 587 295 745]{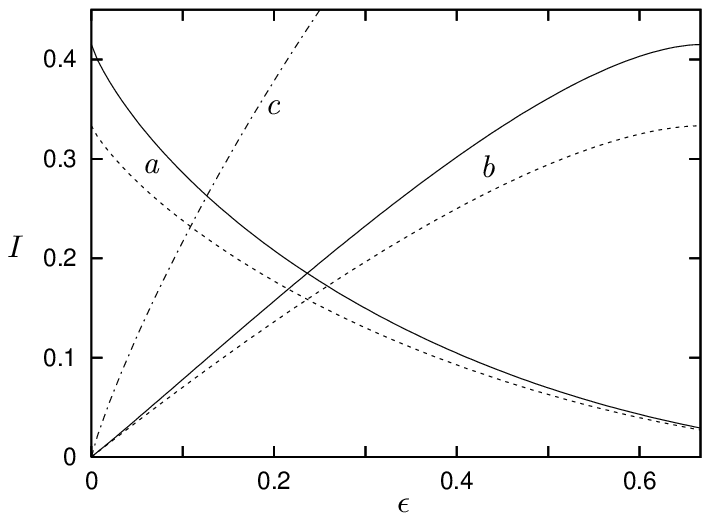}}
\centerline{\includegraphics[bb=80 593 295 750]{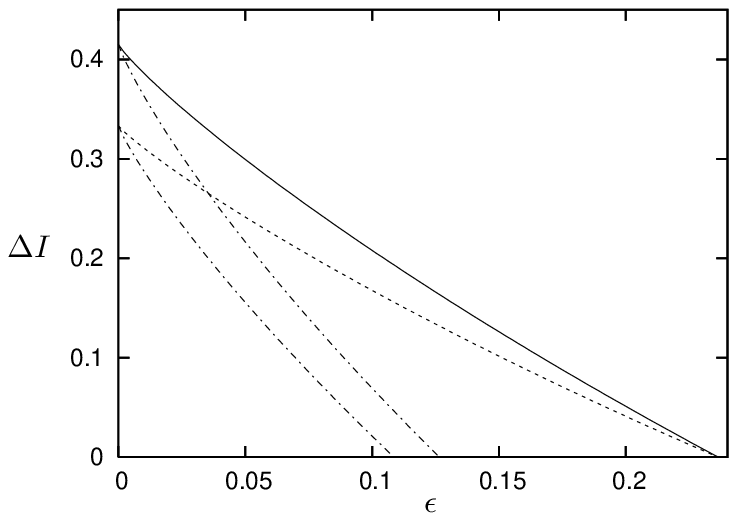}}
\caption{\label{fig:raw}%
Mutual information and yield for key generation by one-way communication.
Top figure: Mutual information about the raw data between Alice and Bob
($I_\mathrm{AB}$, curves \textit{a}) and between Alice and Eve 
($I_\mathrm{AE}$, curves \textit{b}), for the range 
$0\leq\epsilon\leq\frac{2}{3}$ of nonseparable source states.
The solid lines are for the four-outcome POM of the MQT protocols, 
the dashed lines for the six-outcome POM of the six-state protocol.
For both POMs, the intersection of the respective \textit{a} and \textit{b}
curves is at $\epsilon=0.2363$. 
The dash-dotted curve \textit{c} is the common Holevo bound.
It intersects the curves \textit{a} at $\epsilon=0.1265$ and $\epsilon=0.1086$
for the four-outcome POM and the six-outcome POM, respectively.
Bottom figure: The  Csisz\'ar--K\"orner yields 
$\Delta I=I_\mathrm{AB}-I_\mathrm{AE}$ for the mixed-state attack of 
Sec.~\ref{sec:raw} for the four-state POM (solid line) and the six-state POM
(dashed line), for $0\leq\epsilon\leq0.24$.
The two dash-dotted lines show the respective yields for the message attack of
Sec.~\ref{sec:message1way}. 
Note that these plots refer to one-way communication schemes, whereas the
Singapore protocol and the six-state protocol rely on two-way communication.
}
\end{figure}

The resulting mutual information between Alice and Eve, $I_\mathrm{AE}$, and
also that between Bob and Eve, is given by the right-hand side of \Eqref{Iab}
with $\epsilon\to\eta$. 
We have $\epsilon<\eta$ and therefore $I_\mathrm{AB}>I_\mathrm{AE}$ 
for 
\begin{equation}
  \label{eq:CKthr}
\epsilon<\frac{1}{\frac{5}{2}+\sqrt{3}}=0.2363\,.  
\end{equation}
The Csisz\'ar--K\"orner (CK) theorem \cite{CK} then ensures that a secret key
can be generated, by a fitting code for one-way communication, with an
efficiency of $I_\mathrm{AB}-I_\mathrm{AE}$.
We note that the six-state protocol has the same CK threshold \cite{AccInf},
and the CK efficiency of MQT protocols, such as the Singapore protocol, 
is larger for all values below their common CK threshold. 
See Fig.~\ref{fig:raw} for a graphical summary of these observations about the
mixed-state attack.

\subsection{Attacks on the generated key bits}\label{sec:coll+mess}
This CK threshold does not apply, however, if Alice and Bob 
generate the binary key with the iterative procedures of
Secs.~\ref{sec:iteration} and \ref{sec:hybrid},
which involve two-way communication.
The secondary sequences formed in step~2b, from primary sequences
characterized by $\epsilon$, have statistical properties
of the kind stated by the joint probabilities in \Eqref{jointprobs} 
but with $\epsilon$ replaced by $\epsilon^2/[1+\frac{1}{3}(1-\epsilon)^2]$, 
i.e., with quadratically less noise.   
Accordingly, 
the raw key bits that Alice and Bob obtain in the successive
rounds have error probabilities of 
\begin{equation}
  \label{eq:biterror}
  \left[1+\left(\frac{4-\epsilon}{3\epsilon}\right)^{2^{n-1}}\right]^{-1}
\propto{\epsilon}^{2^{n-1}}
\quad\mbox{for the $n$th round},
\end{equation}
so that the quality of the bits improves markedly from one iteration round to
the next.
But also Eve's eavesdropping gets more efficient.
She has two ancillas to examine for each first-round bit, four ancillas for
each second-round bit, \dots, $2^n$ ancillas for each $n$th-round bit. 

We illustrate this matter by a closer look at the first-round bits; 
see the $n=1,L=2$ row in Table~\ref{tbl:thresholds}.
For each of Alice's bits, there is now a two-ancilla state of rank~9, 
conditioned on her bit value~\cite{CohAtt}.
It is as if she were sending a random sequence of two different two-ancilla
states to Eve, both occurring equally frequently.
There are then various strategies for Eve, of which the two extreme ones are
the raw-data attack and the message attack introduced above, and the
collective attack is intermediate.

In the \emph{raw-data attack} \cite{RawDataAtt}, Eve measures the ancillas 
one-by-one and employs the POM of Sec.~\ref{sec:raw} for this purpose, so
that she is not modifying her POM in the light of the information exchanged
by Alice and Bob publicly. 
Just like it is the case for the mixed-state attack, there is no need for Eve
to wait until this information is available. 
She can measure the ancillas immediately after sending the qubits
to Alice and Bob or, equivalently, she could send suitably prepared mixed
two-qubit states to them. 
The public information is only taken into account for the interpretation
of the data. 
While this raw-data attack is rather weak, it is of quite some interest
because it is the best an eavesdropper can do with present-day
technology. 

Not unexpectedly, the \emph{collective attack} \cite{CohAtt} is much stronger. 
For its implementation, Eve applies the POM that is optimized for the
two-ancilla states that she receives. 
The difference is indeed striking: 
For the raw-data attack, Alice and Bob are on the safe side if
$\epsilon<0.3324$, whereas they fall prey to the collective attack
unless $\epsilon<0.2628$. 

We note that both values are above the CK threshold of \Eqref{CKthr}.
If they only have to fear these collective attacks,
Alice and Bob can, therefore, extract secure key bits from their
sequences of raw first-round bits if $\epsilon<0.2628$; and for
$\epsilon<0.2945$ the final-pairing bits of the first round can be trusted 
for key extraction. 
For this purpose, Alice and Bob will have to use an appropriate
error-correcting code and a privacy amplification protocol 
for the effective binary channel thus defined, much like
one does it in other protocols for quantum key distribution. 

But, further information becomes available to Eve during the error correction
and the privacy amplification. 
The \emph{message attack} \cite{CohAtt} takes all of this additional
information into account as well.
As in the collective attack, Eve exploits the conditioned two-ancilla states,
yet in a much more exhaustive manner.
The resulting threshold values are obtained by calculating the
Holevo bound \cite{Holevo,DandW}, 
which in fact is technically much less demanding than the search for the
optimal POMs needed for the raw-data attack and the collective attack.

It follows that the absolute noise thresholds for the first-round keys are
$\epsilon=0.2182$ for the iteration bits and $\epsilon=0.2422$ for the
final-paring bits. 
Both are noticeably below the thresholds for the collective attack, and the
first one is even below the CK threshold.

\begin{table}[t]
\begin{ruledtabular}
\caption{\label{tbl:thresholds}%
Noise thresholds for the Singapore protocol.
For $n=1$, $2$, or $3$ iterations, involving $L=2^n$ letters respectively, 
we report the thresholds for the three eavesdropping attacks considered in 
Sec.~\ref{sec:coll+mess}: the raw-data attack \cite{RawDataAtt}, 
the collective attack \cite{CohAtt}, and the message attack \cite{CohAtt}.
For each attack, there are two columns of threshold values, the thresholds
for the key bits gained during the iteration step of Sec.~\ref{sec:iteration} 
($\epsilon^{\ }_\mathrm{it}$), and the thresholds for the key bits generated
by the final pairing of Sec.~\ref{sec:hybrid} ($\epsilon^{\ }_\mathrm{FP}$) 
if the iteration stops with the $n$th round.
The entries for $L=1$ refer to Renes's original pairing procedure of
Ref.~\cite{RenesTetra}, discussed in Sec.~\ref{sec:Renes}.
The last row states the asymptotic thresholds for ${n,L\gg1}$.
The values for the message attack are absolute thresholds; there is no
eavesdropping strategy with lower thresholds.}
\newlength{\DoubleColWidth}\settowidth{\DoubleColWidth}{collective~attack}
\begin{tabular}{@{}rrllllll@{}}
&& \multicolumn{2}{l}{\makebox[\DoubleColWidth][c]{raw-data attack}}
 & \multicolumn{2}{l}{collective attack}
 & \multicolumn{2}{l}{\makebox[\DoubleColWidth][c]{message attack}}
\\
$n$ & $L$ & \multicolumn{1}{c}{$\epsilon^{\ }_\mathrm{it}$} 
          & \multicolumn{1}{c}{$\epsilon^{\ }_\mathrm{FP}$} 
          & \multicolumn{1}{c}{$\epsilon^{\ }_\mathrm{it}$} 
          & \multicolumn{1}{c}{$\epsilon^{\ }_\mathrm{FP}$} 
          & \multicolumn{1}{c}{$\epsilon^{\ }_\mathrm{it}$} 
          & \multicolumn{1}{c}{$\epsilon^{\ }_\mathrm{FP}$} \\
\hline\rule{0pt}{10pt}%
  & 1 &        & 0.2868 &        & 0.2347 &        & 0.1920 \\
1 & 2 & 0.3324 & 0.3598 & 0.2628 & 0.2945 & 0.2182 & 0.2422 \\
2 & 4 & 0.3742 & 0.4241 & 0.2959 & 0.3405 & 0.2482 & 0.2976 \\
3 & 8 & 0.4143 & 0.4745 & 0.3401 & 0.3649 & 0.2997 & 0.3340 \\
\multicolumn{2}{c}{$\infty$} 
& 0.5091 & 0.5714 & 0.3753 & 0.3893  & 0.3753 & 0.3893 
\end{tabular}
\end{ruledtabular}
\end{table}

The analysis for the second-round bits is analogous but
now there are four ancillas to be examined jointly; 
and there are eight ancillas for each key bit of the third round; and so forth.
We report in Table~\ref{tbl:thresholds} the thresholds found in
Ref.~\cite{RawDataAtt} for the raw-data attack and in Ref.~\cite{CohAtt} for
the collective attack and the message attack, for both the key bits gained
during the iteration step of Sec.~\ref{sec:iteration} and by the final pairing
of Sec.~\ref{sec:hybrid}.
Note that the thresholds are consistently higher for the later
iteration rounds, and the message-attack asymptotic threshold for the
final-pairing bits at $\epsilon=0.3893$ identifies the noise level below which
secure key generation is assuredly possible.
In practice, of course, one would operate well below this value to have a
reasonable efficiency.

Consider, for example, a noise level of $\epsilon=0.25$, slightly above the
BB84 threshold  \cite{BB84-threshold,QBER}. 
If Alice and Bob only have to fear raw-data attacks, all iteration rounds give
secure bits including the $L=1$ final-pairing bits of Renes's scheme.
These $L=1$ bits are not secure, however, if one must protect against
collective attacks.
And if Alice and Bob are afraid of message attacks, they can only use the
iteration bits from the third and later rounds and the final pairing bits from
the second round onwards.

\subsection{Message attack for one-way key generation}
\label{sec:message1way}
If Eve can store the ancillas and process them at a late time, she can also
implement a more powerful attack on one-way key generation than the
mixed-state attack of Sec.~\ref{sec:raw}.
She would wait until Alice and Bob have encrypted a message after
performing the necessary error correction and privacy amplification, 
and would then attack that message.
Thereby, Eve's power is only limited by the Holevo bound.

Now, the six ancilla states conditioned on Alice's measurement result of the
six-state POM and the four ancilla states for the four-state POM are unitarily
equivalent rank-2 states with nonzero eigenvalues $1-\frac{1}{2}\epsilon$ and
$\frac{1}{2}\epsilon$; and the unconditioned ancilla state is the same for
all POMs, the singlet with an admixture of unbiased noise of
(\ref{eq:noisysinglet}). 
Accordingly, there is a common Holevo bound for the four-outcome POM of the MQT
protocols and the six-outcome POM of the six-state protocol; see
Fig.~\ref{fig:raw}.

The corresponding noise thresholds turn out to be quite low: 
${\epsilon=0.1086}$ for the
six-outcome POM, and ${\epsilon=0.1265}$ for the four-outcome POM. 
One is clearly rewarded here for employing a minimal tomographic POM rather
than a POM with redundancy.

\begin{figure}%
\begin{tabular}{r}\rule{0.pt}{158pt}%
\includegraphics[bb=55 585 293 737]{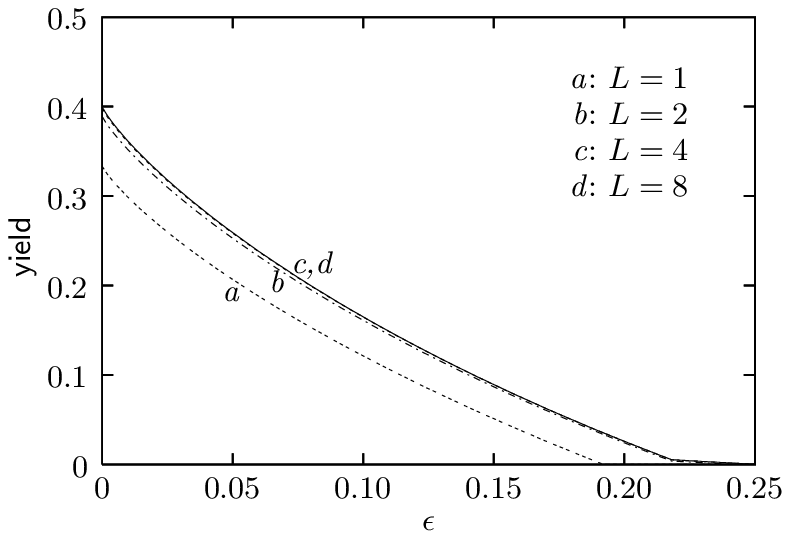}\\[1ex]
\includegraphics[bb=55 634 293 736]{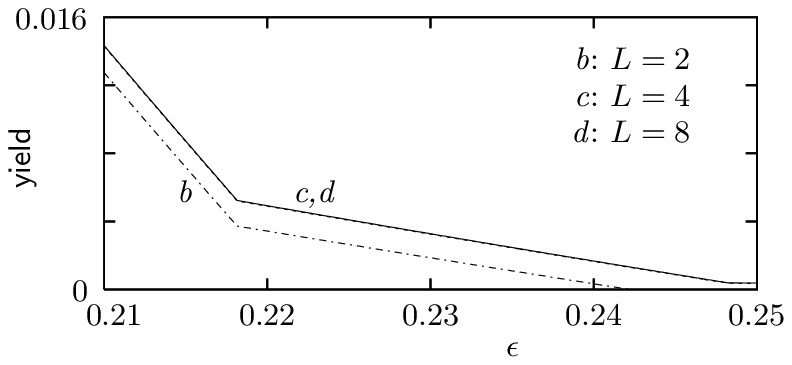}\\[1ex]    
\includegraphics[bb=55 634 293 736]{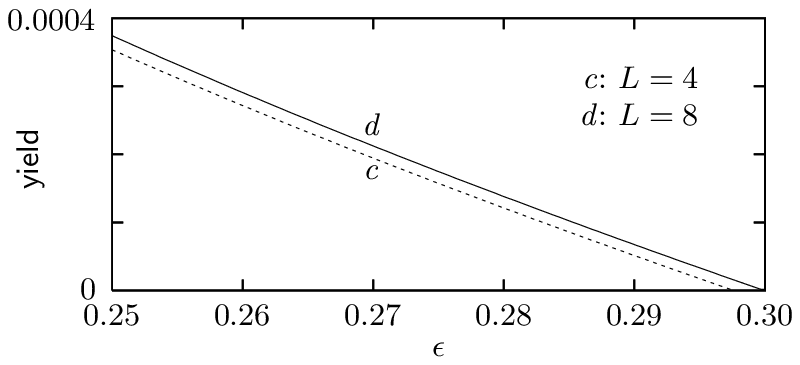}\\[1ex]    
\includegraphics[bb=55 634 293 737]{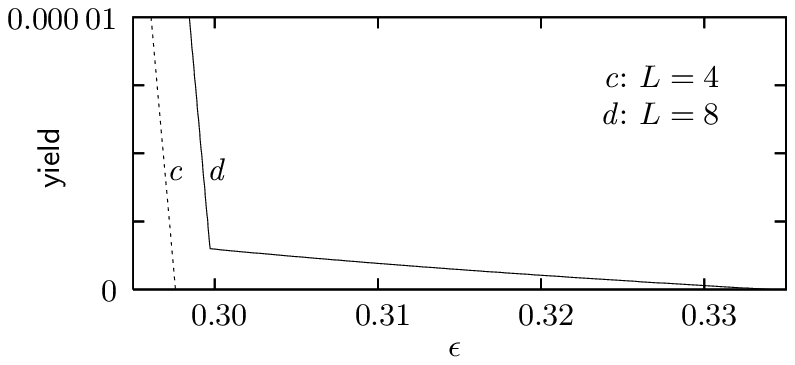}  
\end{tabular}
\caption{\label{fig:yield}%
Yield of the Singapore protocol under the message attack.
The plots show the yield for key generation that uses up to $L=8$ letters of
the raw-data sequences. 
The various curves are for 
$L=1$: Renes's original pairing scheme, 
$L=2$: termination after the first round, 
$L=4$: termination after the second round, and 
$L=8$: termination after the third round.
In the two top plots, which cover the ranges ${\epsilon<0.25}$
and ${0.21<\epsilon<0.25}$, the curves for $L=4$ and $L=8$ are
indiscernible.
They are resolved in the two bottom plots, where ${0.25<\epsilon<0.3}$ and
${0.295<\epsilon<0.335}$. 
The various cusps in the curves are at the threshold values of the last column
pair in Table~\ref{tbl:thresholds}. 
}
\end{figure}

\section{Summary and concluding remarks}
In summary, the Singapore protocol for quantum key distribution introduced
here can tolerate much higher noise levels than the BB84 protocol and is
substantially more efficient than its tomographic competitor, the six-state
protocol. 
The raw data, which consists of two random four-letter sequences with
correlations between them, can be converted into secret key bits by an
iterative procedure that uses two-way communication.  

The results of the security analysis---detailed accounts given in
Refs.~\cite{RawDataAtt,CohAtt}---are concisely summarized in
Table~\ref{tbl:thresholds} and Fig.~\ref{fig:yield}. 
Table~\ref{tbl:thresholds} reports the noise thresholds for the hierarchy of
eavesdropping attacks considered: the raw-data attack, the collective
attack, and the message attack. 
The weakest of them is the raw-data attack, which is the one that is to be
feared within the limitations of today's technology, 
and the most powerful attack is the message attack, which is as strong as the
laws of physics permit and thus determines the absolute noise thresholds. 

These thresholds are clearly visible in the plots of the yield that are shown
in Fig.~\ref{fig:yield}. 
Consistent with the thresholds in Table~\ref{tbl:thresholds}, a nonzero yield
can be had for $\epsilon<0.242$ with a single round of key generation, whereas
two or more rounds are needed for larger noise levels, with three rounds being
sufficient for $\epsilon<0.334$.

It should be kept in mind that our analysis refers to the specified procedure
for generating the key bits that is introduced in Sec.~\ref{sec:2ways}.
When a different procedure is adopted, such as those alluded to in
Ref.~\cite{GetAll}, the noise thresholds will be different as well.

We wish to emphasize that the higher efficiency of the Singapore protocol
does not result from a deliberate asymmetry, as it is the case in the
asymmetric variant of BB84 that is thoroughly studied by Lo, Chau, and
Ardehali in Ref.~\cite{asymmetric}.
Both the six-state protocol and the Singapore protocol are \emph{symmetric} 
in the sense that all of Alice's and Bob's detectors have the same a priori
probability of registering the next qubit.

Asymmetric versions of the Singapore protocol can be realized either by
distorting the tetrahedrons \cite{DistTetra} or by using a source that emits
the qubit pairs in a biased state rather than the singlet. 
In the absence of noise, the efficiency of distorted-tetrahedron schemes can
be as close to unity as one wishes, quite analogously to the asymmetric
BB84-type protocol of Ref.~\cite{asymmetric}, but then it is very costly to
perform reliable tomography. 


\section*{Acknowledgments}
We thank Artur Ekert, Lim Han Chuen, Shiang Yong Looi, Syed Md.\ Assad, 
Jun Suzuki, and Renato Renner for insightful discussions.
H.~K.~Ng would like to thank the Defence Science \& Technology Agency
(DSTA) of Singapore for their financial support.
J.~\v{R}. wishes to thank for the kind hospitality during his visits to
Singapore. 
J.~A. gratefully acknowledges the financial support by the 
Daimler Benz Stiftung.
This work was supported by A$^*$Star Grant 012-104-0040,
by NUS Grant R-144-000-109-112, and by Project SECOQC (FP6-506813) 
of the EU.
Centre for Quantum Technologies is a Research Centre of Excellence funded by
Ministry of Education and National Research Foundation of Singapore.



\begin{thebibliography}{99}
\bibitem{6IAB}
For the six-state protocol under ideal circumstances, one has
six times $p_{kl}=0$, six times $p_{kl}=2p_{k\cdot}p_{\cdot l}=\frac{1}{18}$,
and 24 times $p_{kl}=p_{k\cdot}p_{\cdot l}$.
\bibitem{benchmark}
We are not aware of any efficient error-correcting codes for the six-state 
protocol, so this benchmark is of a theoretical, not of a practical nature.
\bibitem{GetAll}
With more complicated, and thus less practical, procedures one should be able
to get even closer to $0.415$.
We are not aware, however, of any explicit procedure that gives a better yield
than $0.4$.
\bibitem{OtherSources}
The class of acceptable sources could, of course, be larger. 
For the sake of simplicity, we impose here the same criterion as 
in the tomographic protocols of Ref.~\cite{TomoCrypt}.
In practice, Alice and Bob employ twirling to make sure that their joint
probabilities are of the form (\ref{eq:noisy_jointprobs}).
\bibitem{QBER}
The quantum bit error rate of BB84 is $\frac{1}{2}\epsilon$.
\bibitem{DistTetra}
POMs that use distorted, but not degenerate, tetrahedrons are suitable for
qubit tomography \cite{OneLoop,NoLoop}, but they are inferior to those 
using undistorted tetrahedrons \cite{MQT}.
\end{thebibliography}
\end{document}